\documentclass[twocolumn]{aastex631} 

\newcommand\aastex{AAS\TeX}

\newcommand{\cii}{[CII]$_{158\mu m}\,$}
\newcommand{\oiii}{[OIII]$_{88\mu m}\,$}

\newcommand{\oojwst}{[OIII]$\lambda\lambda$4959,5007\AA}
\newcommand{\Hb}{H$\mathrm{\beta}$}
\newcommand{\ciii}{CIII]$\lambda\lambda$1907,9\AA}

\usepackage{gensymb}

\begin{document}

\title{Deep Constraints on \cii in JADES-GS-z14-0:\\Further Evidence for a Galaxy with Low Gas Content at z=14.2}

\author[0000-0001-9746-0924]{Sander Schouws}
\affil{Leiden Observatory, Leiden University, NL-2300 RA Leiden, Netherlands}

\author[0000-0002-4989-2471]{Rychard J. Bouwens}
\affil{Leiden Observatory, Leiden University, NL-2300 RA Leiden, Netherlands}

\author[0000-0002-4205-9567]{Hiddo Algera}
\affil{Institute of Astronomy and Astrophysics, Academia Sinica, 11F of Astronomy-Mathematics Building, No.1, Sec. 4, Roosevelt Rd, Taipei 106216, Taiwan, R.O.C.}

\author[0000-0001-7768-5309]{Renske Smit}
\affil{Astrophysics Research Institute, Liverpool John Moores University, 146 Brownlow Hill, Liverpool L3 5RF, United Kingdom}

\author[0000-0002-5320-2568]{Nimisha Kumari}
\affil{AURA for European Space Agency, Space Telescope Science Institute, 3700 San Martin Drive, Baltimore, MD 21218, USA}

\author[0009-0009-2671-4160]{Lucie E. Rowland}
\affil{Leiden Observatory, Leiden University, NL-2300 RA Leiden, Netherlands}

\author[0009-0005-6803-6805]{Ivana van Leeuwen}
\affil{Leiden Observatory, Leiden University, NL-2300 RA Leiden, Netherlands}

\author[0000-0002-2906-2200]{Laura Sommovigo}
\affil{Center for Computational Astrophysics, Flatiron Institute, 162 Fifth Avenue, New York, NY 10010, USA}

\author[0000-0002-9400-7312]{Andrea Ferrara}
\affil{Scuola Normale Superiore, Piazza dei Cavalieri 7, 50126 Pisa, Italy}

\author[0000-0001-5851-6649]{Pascal A. Oesch}
\affil{Departement d’Astronomie, Universit\'e de Gen\'eeve, 51 Ch. des Maillettes, CH-1290 Versoix, Switzerland}
\affil{Cosmic Dawn Center (DAWN), Copenhagen, Denmark}
\affil{Niels Bohr Institute, University of Copenhagen, Jagtvej 128, DK-2200 Copenhagen N, Denmark}

\author[0000-0003-2000-3420]{Katherine Ormerod}
\affil{Astrophysics Research Institute, Liverpool John Moores University, 146 Brownlow Hill, Liverpool L3 5RF, United Kingdom}

\author[0000-0001-7768-5309]{Mauro Stefanon}
\affil{Leiden Observatory, Leiden University, NL-2300 RA Leiden, Netherlands}

\author[0000-0003-2164-7949]{Thomas Herard-Demanche} 
\affil{Leiden Observatory, Leiden University, NL-2300 RA Leiden, Netherlands}

\author[0000-0001-6586-8845]{Jacqueline Hodge}
\affil{Leiden Observatory, Leiden University, NL-2300 RA Leiden, Netherlands}

\author[0000-0001-7440-8832]{Yoshinobu Fudamoto} 
\affiliation{Center for Frontier Science, Chiba University, 1-33 Yayoi-cho, Inage-ku, Chiba 263-8522, Japan}

\author[0000-0001-8887-2257]{Huub R\"ottgering}
\affil{Leiden Observatory, Leiden University, NL-2300 RA Leiden, Netherlands}

\author[0000-0002-4389-832X]{Paul van der Werf}
\affil{Leiden Observatory, Leiden University, NL-2300 RA Leiden, Netherlands}

\begin{abstract}

We present deep ALMA observations targeting the \cii line in JADES-GS-z14-0, the most distant known galaxy at z=14.1793. We do not detect the \cii line in our deep observations, implying a luminosity of $<$6$\times10^7$ L$_{\odot}$ (3$\sigma$) for the target. Comparing this with the detected \oiii line, we constrain the [OIII]/[CII] ratio to be $>$3.5, significantly improving our probe of the ionization parameter $U$.  The observed ratio is higher than analogues in the local universe, but consistent with galaxies at $z\approx6$-9.  Through ISM modeling, we infer extreme ionizing conditions with log(U)$>-$2.0, likely requiring a young stellar population. Our modeling also indicates a relatively low gas density ($51_{-32}^{+116}$ cm$^{-3}$), significantly lower than expected from lower redshift trends. We infer a relatively high gas-phase metallicity (16$\pm$6\% solar) consistent with previous results and implying a rapid build-up of metals.  Finally, using [CII] as a molecular/cold gas mass tracer, we infer a low gas fraction ($f_\mathrm{gas} < 0.77$), consistent with previous estimates of the dynamical mass from \oiii.  Combined with the low observed gas density, lack of dust and high ionization parameter, this suggests strong feedback processes are playing an important role in the evolution of this galaxy. Our observations show that JADES-GS-z14-0 is a rapidly evolving galaxy with extreme ISM conditions, shedding light on the earliest phases of galaxy formation.

\end{abstract}



\section{Introduction} \label{sec:introduction}

The discovery and spectroscopic confirmation of galaxies at $z>10$ has recently become possible due to the groundbreaking capabilities offered by the James Webb Space Telescope (\textit{JWST}) (e.g., \citealt{Curtis-Lake_2022,zavala2024}). While initial follow-up with the Atacama Large Millimeter/submillimeter Array (ALMA) of bright {\it JWST} targets was largely unsuccessful (e.g. \citealt{bakx2023,Popping_2023,Yoon_2023,fujimoto2023_highzalma,kaasinen2023}), this changed with the detection of \oiii in JADES-GS-z14-0 (hereafter GS-z14) at $z=14.1793\pm$0.0007 \citep{carniani2024_oiii,schouws2024} and GHZ2/GLASSz12 at $z=12.3327\pm$0.0035 \citep{zavala2024_alma}. This has provided a new and unique view of these galaxies, revealing a rapid build-up of metals, relatively low densities and potentially a low gas fraction \citep{carniani2024_oiii,schouws2024,zavala2024_alma} and has opened the door for further studies of $z>10$ galaxies with ALMA.

\begin{figure*}[th!]
\epsscale{1.15}
\plotone{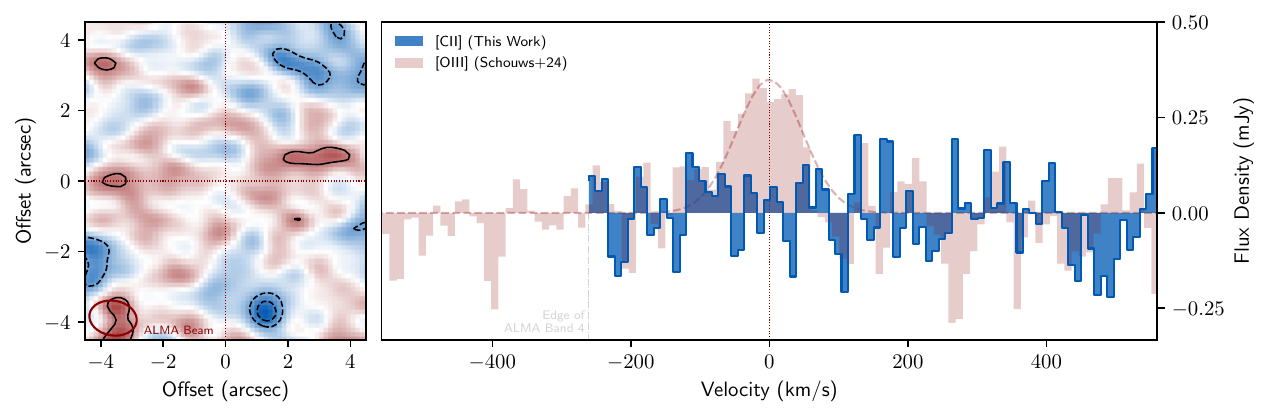} 
\caption{Sensitive constraints just obtained on \cii emission from GS-z14 with ALMA (2024.A.00007.S: PI Schouws). \textit{Left Panel:} Moment-0 map collapsed over 130 km/s at the expected frequency of \cii.  To provide a sense of the noise levels, a 10$"$$\times$10$"$ area is presented, with contours drawn at the $-$3, $-$2, 2 and 3-$\sigma$ levels, where dashed contours are used for the negative significance levels.  The ALMA beam-size is shown as a red ellipse in the lower left corner.  The position of GS-z14 is highlighted by the red cross-hairs at the center. \textit{Right Panel:} Spectrum extracted within a 0.5\arcsec aperture centered on the position of GS-z14 (\textit{blue bars}).  A small amount of smoothing (bin size of 2-spectral elements) is applied.  The spectral coverage is limited due to its proximity of \cii to the edge of ALMA Band 4 at 125 GHz. For context, emission from \oiii line is also using a similar aperture and extraction scheme (\textit{red bars}). The red-dashed line shows the \citet{schouws2024} fit to the \oiii line.  It is clear that \cii is not detected. (We note that equal flux in both lines would result in a ratio [OIII]/[CII]$\sim$1.8 due to the dependence of the derived luminosity on frequency, e.g., \citealt{Solomon_1992}).} \label{fig:spec}
\end{figure*}

These developments have provided a glimpse into the physical conditions of the interstellar medium (ISM) for the earliest galaxies, which has been a longstanding goal in high redshift astronomy \citep[e.g.,][]{Robertson2022}.   Emission lines provide us with key information to constrain these physical conditions.  The advent of {\it JWST} has revolutionized the detection of rest-frame optical emission lines that have traditionally been essential to characterize the ISM. However, this becomes much more challenging at $z>9.5$ when strong rest-frame optical emission lines like \oojwst$\,$and \Hb\,move out of the spectral coverage of NIRSpec.  While these lines can still be targeted by MIRI \citep{zavala2024,AlvarezMarquez2024}, the cost is much higher due to the lower sensitivity of MIRI and lack of multiplexing.  In this regime, ALMA can provide competitive and complementary insights that rival or exceed those possible with {\it JWST} \citep{schouws2024,carniani2024_oiii}.

The aim of this study is to report on new deep \cii observations of GS-z14, currently the most distant spectroscopically-confirmed galaxy known.  GS-z14 has measurements of \oiii, \ciii, and a photometric constraint on \oojwst + \Hb\, \citep{carniani2024,carniani2024_oiii,schouws2024,helton2024}. Although this makes GS-z14 one of the best characterized $z>10$ galaxies, this suite of emission lines is insufficient for an unambiguous characterization of its ISM, with sizable uncertainties in key parameters, particularly the ionization parameter \citep[][]{carniani2024_oiii, schouws2024}.  Fortunately, with sensitive observations of \cii we can substantially reduce many existing uncertainties in our modeling of the ISM conditions of GS-z14. 

Of particular relevance is the \oiii/\cii ratio (hereafter [OIII]/[CII]), which provides us with invaluable constraints on the ionization conditions in the ISM \citep[e.g.][]{arata2020,harikane2022,katz2022}. This ratio also has been a key focus of study both in the local universe and galaxies at $z\sim 6$-9, facilitating direct comparisons to our observational results \citep{inoue2016,hashimoto2019,laporte2019,bakx2020,carniani2020,harikane2020,akins2022,Witstok_2022,algera2024,bakx2024,fujimoto2024_s04950}.

Throughout this paper we assume a standard $\mathrm{\Lambda}$CDM cosmology with $H_0=70$ km s$^{-1}$ Mpc$^{-1}$, $\Omega_m=0.3$ and $\Omega_{\Lambda}=0.7$. Magnitudes are presented in the AB system \citep{oke_gunn_1983ApJ...266..713O}. For star formation rates (SFR) and stellar masses, we adopt a Chabrier initial mass function \citep{chabrier2003}. Error-bars indicate the $68\%$ confidence interval unless specified otherwise. All measured and derived physical quantities are corrected for gravitational lensing by a factor of 1.17$\times$ \citep{carniani2024}. Logarithms use base 10 unless specified otherwise.

\section{Observations and Data-Reduction} \label{sec:observations}

\subsection{Previous Observations}  \label{sec:previous-observations}

GS-z14 is the most distant galaxy currently known and simultaneously the second most luminous $z>8$ galaxy with a spectroscopic redshift (only GN-z11 is more luminous by a factor $\sim$2$\times$: \citealt{Oesch_2016,bunker2023}). In contrast to other bright $z>10$ galaxies (e.g., GN-z11 or GHZ2: \citealt{castellano2024}), GS-z14 has a relatively extended morphology (FWHM$\sim$0.16\arcsec), indicating that its luminosity is primarily driven by star formation as opposed to an active galactic nucleus (AGN).

It was initially discovered in deep imaging of the GOODS-South field obtained by the {\it JWST} Advanced Deep Extragalactic Survey \citep[JADES;][]{Eisenstein_2023}, the First Reionization Epoch Spectroscopic COmplete Survey \citep[FRESCO;][]{Oesch_2023}, and JADES Origins Field (JOF) program \citep{Eisenstein_2023_JOF} and subsequently targeted for follow-up spectroscopy with NIRSpec.

\begin{figure}[t]
\epsscale{1.175}
\plotone{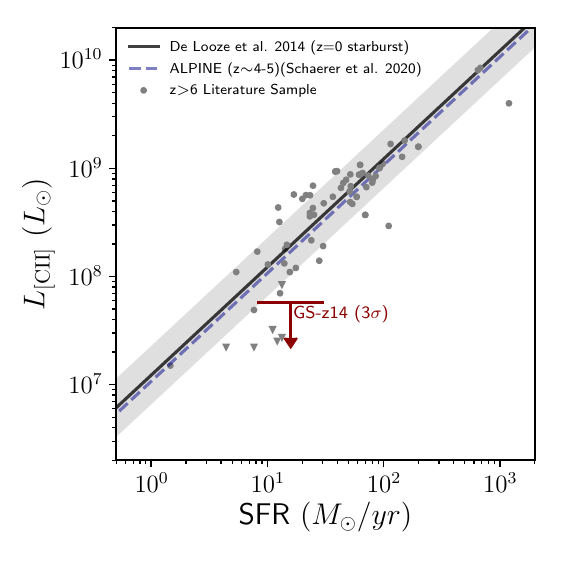} 
\caption{Comparison of the limiting \cii luminosity (\textit{downward arrow}) inferred from GS-z14 against its SFR.  The horizontal line associated with the upper limit extends over the full range in SFR derived for GS-z14 from \citet{carniani2024} and \citet{helton2024}.  Previous studies have revealed a tight correlation between the SFR and L$_{\mathrm{[CII]}}$ in local galaxies \citep{delooze2014} and at high-redshift \citep[e.g.][Schouws et al. in prep.]{carniani2017, schaerer2020}  GS-z14 lies below the relation over the full range of SFRs.} \label{fig:lcii_sfr}
\end{figure}

The NIRSpec spectroscopy showed a strong break at $\sim$1.85$\mu$m, consistent with a Lyman break at $z\sim14$, but also a lack of strong rest-frame UV emission lines, resulting in considerable uncertainty in the spectroscopic redshift estimate (z = 14.32$^{+0.08}_{-0.20}$) \citep{carniani2024}.

GS-z14 was subsequently targeted in an ALMA DDT programme (\#2023.A.00037.S: PI Schouws) to scan for the luminous \oiii line.  These observations were successful, resulting in a 6.6$\sigma$ detection of the \oiii line at z=14.1793$\pm$0.0007 in only 2.3 hours on-source integration \citep{carniani2024_oiii,schouws2024}.  The discovered \oiii line detection further confirmed the reality of the tentative 3.7$\sigma$ \ciii\, line candidate that was previously identified in the NIRSpec spectroscopy.

Finally, GS-z14 is also detected by MIRI in the F770W filter \citep{helton2024}. This filter covers the rest-frame optical emission of GS-z14, corresponding to wavelengths of 4.4 to 5.7$\mu$m, enabling a constraint on the luminosity of the strong \oojwst\, and \Hb\, emission lines.

\subsection{ALMA \cii Follow-up}  \label{sec:description-of-observations}
The observations presented in this work were obtained as part of a Cycle 11 Director's Discretionary Time (DDT) program (\#2024.A.00007.S, PI: Schouws). We target the \cii line, which is expected at 125.20 GHz and falls just within ALMA Band 4 (i.e. within 0.2 GHz from the edge of the band). We use a standard observing set-up with a single tuning and two spectral windows in each of the side-bands. 

The observations were carried out between 11 and 19 January 2025 with an average PWV of 2.38mm (ranging from 1.67mm to 2.98mm), achieving the requested sensitivity in 8.15 hours on-source (12.2 hours including overheads). The target was observed with a moderately compact configuration with baselines ranging from 15m to 783m, resulting in a natural-weighted synthesized beam with a resolution of 1.35$\times$0.99\arcsec\, at the expected frequency of \cii. At this resolution, we expect that all \cii flux from GS-z14 lies within a single beam, preventing signal dilution and loss of S/N \citep[e.g.][]{carniani2020}. The resolution is furthermore similar to the previous observations targeting the \oiii line \citep{schouws2024}.

For the imaging, we use reduced and calibrated measurement sets obtained through the ALMA archival data service. The individual measurement sets are combined and time-averaged in bins of 30 seconds to reduce data-volume and we carefully verify that time-average smearing does not impact fluxes even at the edge of the field of view \citep[e.g.][]{thompson2017}. 

Imaging of the calibrated visibilities was performed with natural weighting using the \textsc{tclean} task in \textsc{Casa} (v6.6.1-17) \citep[\textsc{Casa;}][]{Hunter_2023}. We clean to a depth of 3$\sigma$ using automasking with the recommended settings for low-resolution ALMA data \citep{kepley2020}. At the expected \cii frequency, we generate a moment-0 map including channels within 130 km/s, to approximately match the inferred FWHM of the \oiii line (126 km/s). The resulting moment-0 map has a rms of 7.2 mJy$\cdot$km/s.

We do not detect continuum emission from GS-z14, constraining the flux to $< 9.5\,\mu\mathrm{Jy\,beam}^{-1}$ ($3\sigma$). This is slightly deeper than the previous constraints from the ALMA Band 6 observations ($< 15.1\,\mu\mathrm{Jy\,beam}^{-1}$), but does not improve constraints on the dust content of GS-z14 due to the shape of the dust SED. We therefore refer to \citet{schouws2024} for a more detailed discussion on the dust content of GS-z14.

\begin{figure}[t]
\epsscale{1.175}
\plotone{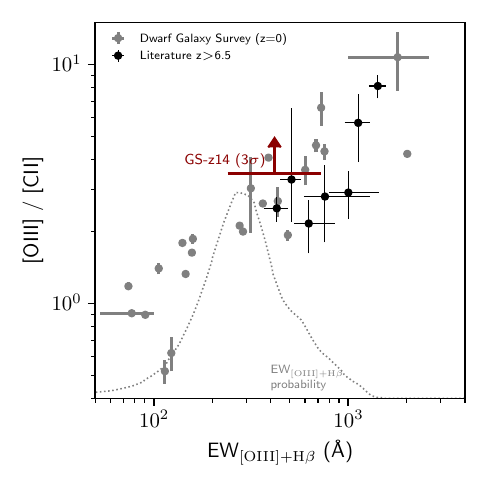} 
\caption{[OIII]/[CII] ratio as a function of [OIII]$_{4959,5007}$+H$\beta$ equivalent width. GS-z14 is shown as a lower limit (maroon), and its EW probability distribution from \citet{helton2024} is overlaid via the dotted line. High-$z$ galaxies from the literature are overplotted as black markers (\citealt{Witstok_2022}; Algera et al.\ in preparation). Moreover, low-redshift galaxies from the Dwarf Galaxy Survey are shown as grey symbols (\citealt{Kumari2024a}).} \label{fig:ew_vs_ratio} 
\end{figure}

\section{Results} \label{sec:results}

\subsection{Non-detection of \cii}  \label{sec:non-detection}

We do not detect significant emission at the location and expected \cii frequency of GS-z14. This is illustrated in Figure \ref{fig:spec}, where we show the moment-0 map and spectrum as extracted within an aperture with a diameter of 0.5\arcsec\, centered on GS-z14.  We do not even find evidence for tentative ($\sim$1-2$\sigma$) \cii emission from GS-z14.  For context, we also include the extracted spectrum of \oiii for GS-z14 using the same method.

Based on the rms of the moment-0 map, we therefore constrain the apparent \cii flux to be below $<$22 mJy$\cdot$km/s (3$\sigma$). This corresponds to a limiting luminosity of $<$6$\times$10$^7$ L$_{\odot}$ (corrected for lensing)\footnote{The impact of the cosmic microwave background (CMB) on \cii (assuming optically thin transmission) can be estimated by $\eta$=F$^{obs}_{\nu}$/F$^{intr}_{\nu}$=$1 - B_{\nu}(T_{CMB}) / B_{\nu}(T_{spin}^{[CII]}$) \citep[e.g.][]{kohandel2019}. The CMB temperature at this redshift is $T_{CMB}=41.4$K and the \cii spin temperature depends strongly on the ISM conditions of the emitting gas. At high redshifts \cii primarily originates from photo-dissociation region (PDR) \citep[e.g.][]{gullberg2015,novak2019,cormier2019}, which have relatively high densities ($log(n_e)\sim2-4$). Given the extreme nature of GS-z14, the kinetic temperature is likely to be high ($T_{kin}\sim500-1000$K). Based on this, we estimate a spin temperature of $T_{spin}^{[CII]}=100-1000$K \citep{gong2012}. This gives $\eta\approx0.60-0.96$, implying a potentially modest suppression of signal from the CMB.  Given the large uncertainty in the physical properties of the \cii emitting gas in GS-z14, the reported \cii luminosities in the paper are not corrected for the impact of the CMB}.

Galaxies in the local and high redshift universe show a moderately tight correlation between their SFRs and \cii luminosities \citep{delooze2014,herrera-camus2015,schaerer2020}. In Figure \ref{fig:lcii_sfr}, we show that the \cii luminosity of GS-z14 lies below this relation. The estimated SFR of GS-z14 ranges from $10_{-2}^{+2}$ M$_{\odot}$/yr \citep{carniani2024} to $25_{-5}^{+6}$ M$_{\odot}$/yr \citep{helton2024}. Depending on the assumed SFR, GS-z14 falls either well below the relation or is marginally consistent, given the observed 0.3dex scatter in the local relation \citep{delooze2014}.  For the lower SFR estimates, the \cii deficit in GS-z14 would be a good match to the deficiency seen in other high redshift galaxies \citep[e.g.][]{harikane2020,Binggeli2021,glazer2024}. This may therefore indicate that the SFR estimate of \citet{helton2024} for GS-z14 is too high.

Leveraging our earlier determination of the \oiii luminosity (2.1$\times$10$^8$ L$_{\odot}$: \citealt{schouws2024}) for GS-z14, we constrain the \oiii to \cii ratio to be $>$3.5. This is consistent with what is typically observed for $z$$\sim$6 - 9 galaxies, which show line ratios of $\mathrm{[OIII]/[CII]} \approx 2 - 10$ \citep[e.g.][]{harikane2020,carniani2020,Witstok_2022,algera2024}. 

ISM modelling has shown that the [OIII]/[CII] ratio correlates with the ionization parameter, which can be elevated by bursts of recent star formation \citep[e.g.][]{ferrara2019, algera2024}. Observationally, recent bursts in star formation result also enhance the equivalent width of \oojwst+\Hb\, (hereafter EW$_{[\mathrm{OIII}]+H\beta}$) due to an increased contrast between enhanced emission lines driven by recent star formation and the underlying stellar continuum.

\begin{figure*}[th!]
\epsscale{1.175}
\plotone{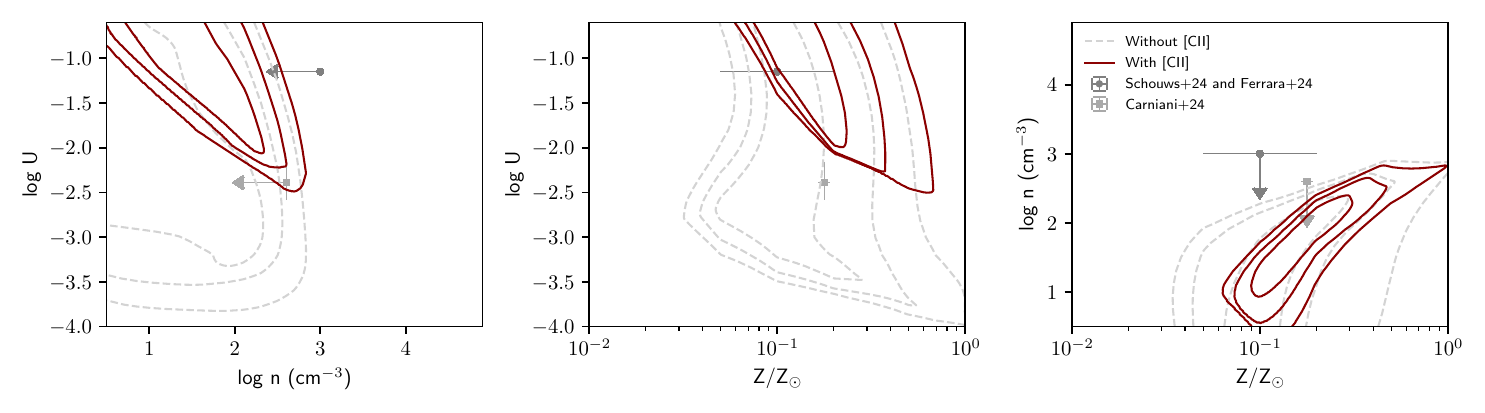} 
\caption{Constraints on the ISM of GS-z14. Each panel shows constraints on two ISM properties while we marginalize over the remaining parameter. We show 1, 2 and 3$\sigma$ constraints for fits both with and without the constraints on \cii (red and gray dashed contours, respectively). The inclusion of \cii significantly improves the constraints on the ISM properties, particularly the ionization parameter. For reference we also indicate previous estimates on the ISM properties of GS-z14 from \citet{carniani2024_oiii} and \citet{schouws2024} (with log(U) estimated by \citealt{Ferrara2024a}).} \label{fig:ISM} 
\end{figure*}

These effects result in a tight correlation between EW$_{[\mathrm{OIII}]+H\beta}$ and [OIII]/[CII] as shown in Figure \ref{fig:ew_vs_ratio}, we also show our constraint on [OIII]/[CII] for GS-z14 in this figure together with the estimate for EW$_{[\mathrm{OIII}]+H\beta}$ from \citet{helton2024} (370$^{+360}_{-130}$ \AA). Our constraint implies that [OIII]/[CII] in GS-z14 is be higher than expected based on its EW$_{[\mathrm{OIII}]+H\beta}$, but could also imply that EW$_{[\mathrm{OIII}]+H\beta}$ might be underestimated by \citet{helton2024}.

We discuss the physical implications of the [OIII]/[CII] ratio of GS-z14 in the context of the known high-redshift galaxy population in the next section.


\section{Discussion} \label{sec:discussion}

\subsection{Constraints on the ISM of GS-z14} \label{sec:ism_constraints}

The detection of line emission from galaxies (or physically meaningful limits) provides us with a powerful way to constrain the ISM conditions in galaxies.  In the case of GS-z14, flux measurements are available for \oiii and \ciii$\,$lines, an upper limit is now available on \cii, and finally  \oojwst + \Hb\ fluxes have constraints thanks to deep MIRI photometry on the source \citep{carniani2024,carniani2024_oiii,helton2024,schouws2024}.

To enable a constraint on the ISM properties of GS-z14 based on these measurements, we use Bayesian inference on an extensive grid of photo-ionization models made with {\sc Cloudy} \citep[v17.02;][]{ferland2017}. In particular, the models consist of an HII region that smoothly transitions into a PDR until a fixed optical depth (A$_V$=10) in a plane parallel geometry. The models have three main parameters to describe the ISM, the ionization parameter ($U$), density ($n$) and metallicity ($Z$). For more details on the models, we refer to \citet{Witstok_2022}. 


In Figure \ref{fig:ISM}, we show the results of our fitting, both with and without our new constraints on the \cii luminosity of GS-z14. The current emission line constraints are well described by models with a high ionization parameter, low density, and moderate metallicity (see Table \ref{table:properties}).

This is consistent with previous modeling efforts of the ISM of high redshift galaxies \citep[e.g.][]{harikane2020,katz2022,nakazato2023}, showing that high ionization parameters are necessary to explain the high observed [OIII]/[CII] ratios at $z > 6$. Our analysis suggests that GS-z14 has a high $\log(U) > -2$, which is consistent with the range of ionization parameters found by \citet{reddy2023} and \citet{tang2023} for a sample of $z\approx3$-9 galaxies observed with {\it JWST}/NIRSpec ($-3 \lesssim \log U \lesssim -1.5$). Moreover, the inferred value for GS-z14 is consistent with the ionization parameter of GHZ2, for which \citet{calabro2024} infer $\log(U) = -1.75 \pm 0.16$. This suggests that high ionization parameters are ubiquitous in the early Universe. 

On the other hand, earlier estimates preferred lower values for the ionization parameter for GS-z14. For instance, \citet{helton2024} measured $\log(U) \approx -2.5$ from NIRCam and MIRI F770W photometry, with the latter band containing the photometric signature of the [OIII]$_{4959,5007}$ and H$\beta$ lines. Similarly, \citet{carniani2024_oiii} inferred a fiducial value of $\log(U) = -2.39_{-0.19}^{+0.23}$ by also including the ALMA \oiii{} detection in GS-z14. The higher $\log(U) > -2$ we infer here highlights the value of our current probe of \cii has for breaking the degeneracy between ionization parameter, density and metallicity.

While a high ionization parameter is a possible driver of the high [OIII]/[CII] ratios of distant galaxies (e.g., \citealt{harikane2020}), other interpretations exist in the literature. For example, a low PDR covering fraction is also expected to result in a high [OIII]/[CII] ratio \citep{harikane2020}, and is qualitatively consistent with the low inferred gas fraction for GS-z14, which may suggest it hosts an outflow (Section \ref{sec:outflows}).


Finally, the production of carbon is thought to lag behind that of oxygen, which is also expected to result in elevated [OIII]/[CII] ratios (e.g., \citealt{katz2022,fujimoto2024_s04950,nyhagen2024}). However, leveraging the \ciii{} detection for GS-z14, \citet{carniani2024_oiii} do not find evidence for an elevated C/O abundance in the galaxy, although the uncertainties on the abundance ratio are large. 




\subsection{Low Gas Content in GS-z14: Evidence for A Post Outflow Phase?} \label{sec:outflows}

In recent years, the \cii{} line has seen frequent use as a molecular gas mass tracer at high redshift (e.g., \citealt{zanella2018,dessauges-zavadsky2020,aravena2024}), owing to the faintness of canonical cold gas tracers such as CO at this epoch. While the precise conversion from \cii{} luminosity to molecular gas mass remains debated (e.g., \citealt{casavecchia2024,algera2025}), we use the commonly adopted value from \citet{zanella2018} of $M_\mathrm{mol}/M_\odot = 31 L_{\text{\cii{}}} / L_\odot$ to place an upper limit on the molecular gas mass of GS-z14 of $\log(M_\mathrm{mol}/M_\odot) < 9.2$ ($3\sigma$).

Adopting the stellar mass from \citet{helton2024} (Table \ref{table:properties}), the \cii{}-based gas mass estimate implies a gas fraction of $f_\mathrm{gas} = M_\mathrm{mol} / (M_\mathrm{mol} + M_\star) < 0.77$ for GS-z14. This is significantly below what is expected for a galaxy with a metallicity of $Z/Z_\odot \sim 0.1 - 0.2$ (c.f., $f_\mathrm{gas} \gtrsim 0.9$; e.g., \citealt{devis2019,algera2025}), which may suggest the gas in GS-z14 is efficiently expelled through outflows. 

The low \cii based gas mass estimate is consistent with the low dynamical mass of GS-z14 based on the line width of the \oiii{} line of $\log(M_\mathrm{dyn}/M_\odot)\sim9.0$ \citep{carniani2024_oiii,schouws2024}. After subtracting the stellar mass, this dynamical estimate would imply a gas fraction of only $\sim$30\% \citep{carniani2024_oiii}.

Further evidence for feedback comes from the lack of dust in GS-z14 as implied by the low A$_V$=0.31 \citep{carniani2024} and non-detection of the dust continuum by ALMA \citep{carniani2024_oiii,schouws2024,algera2025}. Dust enrichment is expected to follow rapidly after the onset of star formation as a consequence of production by Supernovae and for the observed stellar mass, GS-z14 should consequently be detectable with ALMA if all the dust was still present.

Detailed modeling of GS-z14 by \citet{Ferrara2024a} and \citet{Ferrara2024b} suggests that the galaxy recently experienced an outflow, causing its dust to be blown out to large radii. Such an extended dust distribution is subject to a less intense radiation field, enabling it to efficiently cool, and rendering it undetectable at the current depth of the ALMA observations while also explaining the low A$_V$ as observed by {\it JWST}. Following this previous outflow event, \citet{Ferrara2024b} suggest that GS-z14 currently resides in a `post-outflow' phase, characterized by a low current SFR and little obscuration. In other words, the observable properties of GS-z14 appear to be shaped by strong feedback from the galaxy.

Finally, this would also be qualitatively consistent with the high ionization parameter and low density we infer from our observations.  The same star-bursting phase would not only produce a young stellar population and high ionization parameter \citep[e.g.][]{reddy2023}, but also drive strong feedback in the galaxies, expelling or otherwise puffing up its gas reservoir, leaving the source with the low gas density we infer  ($n \sim 50\,\mathrm{cm}^{-3}$: well below the typical densities observed for $z\gtrsim6$ galaxies, e.g., \citealt{isobe2023}).


\vspace{1.0cm}
\begin{table}[ht]
\centering
\caption{Properties of GS-z14}
\vspace{-0.4cm}
\begin{tabular}{p{0.6\columnwidth} p{0.3\columnwidth}}
\hline
\textbf{Parameter} & \textbf{Value} \\ \hline
RA & 03:32:19.9049 \\ 
Dec & $-$27:51:20.265 \\ 
Redshift & $14.1793(7)$ \\ 
M$_{UV}$$^a$ & $-20.81\pm0.16$ \\ 
Stellar Mass($\log(M_\odot)$)$^b$  & $8.7_{-0.4}^{+0.5}$ \\
Star Formation Rate (M$_{\odot}$/yr) & $10_{-2}^{+2}$ $^{(a)}$ ; $25_{-5}^{+6}$ $^{(b)}$ \\
CIII] Luminosity ($10^8\,$L$_{\odot}$)$^{a}$ & 2.0$\pm$0.5$^c$\\ 
$[$OIII]+\Hb\, Luminosity ($10^8\,$L$_{\odot}$)$^{b}$ & 22$\pm$10$^c$ \\
\oiii Luminosity ($10^8\,$L$_{\odot}$)$^d$ & 2.1$\pm$0.5\\
\cii Luminosity ($10^8\,$L$_{\odot}$) & $<$ 0.6$^e$ \\
$[$OIII]/[CII] & $>$ 3.5$^e$ \\
log(U)$^f$ & $>-2.0$$^g$ \\
log(n) (cm$^{-3}$)$^f$ & $1.7_{-0.4}^{+0.5}$ \\
Z/Z$_{\odot}$$^f$ & $0.16_{-0.06}^{+0.06}$ \\
\hline
\end{tabular}
\begin{flushleft} 
\footnotesize{\textit{Notes:} All values have been corrected for a lensing magnification of 1.17$\times$ \citep{carniani2024}. $^a$ Value from \citet{carniani2024}; $^b$ Value from \citet{helton2024}; $^c$ Corrected for extinction; $^d$ Value from \citet{schouws2024}; $^e$ 3$\sigma$ limit; $^f$ Based on the ISM fitting presented in this work; $^g$ 95\% confidence interval. }  \label{table:properties}
\end{flushleft}
\end{table}

\section{Summary} \label{sec:discussion}


We presented new results based on sensitive ALMA observations of the \cii{} line in JADES-GS-z14-0, the most distant galaxy currently known at $z=14.1793\pm0.0007$ \citep{carniani2024,carniani2024_oiii,schouws2024}. Our main conclusions are as follows:

\begin{itemize}
    \item We do not detect the \cii{} line in GS-z14 down to a (lensing-corrected) luminosity of $L_{\text{\cii{}}} < 6\times10^7\,L_\odot$ at $3\sigma$ (Figure~\ref{fig:spec}). This places the galaxy below the \cii{}-SFR relation (Figure~\ref{fig:lcii_sfr}), which is qualitatively consistent with what has been observed for low-SFR galaxies at $z\approx6-9$ \citep{harikane2020,Binggeli2021,glazer2024}. 

    \item Combined with a previous detection of the \oiii{} line in GS-z14 \citep{carniani2024_oiii,schouws2024}, we infer a lower limit on its [OIII]/[CII] ratio of $>3.5$ (Figure~\ref{fig:ew_vs_ratio}). While larger than what is typically seen in galaxies in the local Universe, this elevated line ratio is consistent with that seen in many z$\sim$6-9 galaxies \citep[e.g.][]{inoue2016}. 

    \item Leveraging the sensitive constraints on the \cii{} luminosity, we investigate the ISM conditions of GS-z14 (Figure~\ref{fig:ISM}). In agreement with analyses by \citet{carniani2024,carniani2024_oiii} and our own analysis \citep{schouws2024}, we find that GS-z14 is already significantly metal-enriched, with $Z/Z_\odot = 0.16 \pm 0.06$. Crucially, our new \cii{} observations allow us to break a previously existing degeneracy between the ionization parameter $U$ and density of GS-z14.  We find an ionization parameter of $\log(U) > -2$, higher than what was previously inferred for the galaxy. Moreover, we infer a low density for $\log(n/\mathrm{cm}^{-3})$ of $1.7_{-0.4}^{+0.5}$.

    \item Using \cii{} as a gas mass tracer, we infer a low molecular gas fraction of $f_\mathrm{gas} < 0.77$ for GS-z14, consistent with that obtained via a dynamical estimate from the previously detected \oiii{} line \citep{carniani2024_oiii}.  This is lower than expected for galaxies at a metallicity similar to GS-z14 (e.g., see \citealt{algera2025}), suggesting the efficient removal of its gas through outflows.
\end{itemize}

Overall, the high ionization parameter, low gas density, lack of dust, and low gas fraction of GS-z14 suggest the galaxy is subject to strong feedback effects, likely driven by its star-bursting nature. These extreme physical conditions highlight the complex nature of galaxy formation in the early Universe.

\begin{acknowledgments}
\subsection*{Acknowledgments} 
This paper makes use of the following ALMA data: ADS/JAO.ALMA 2024.A.00007.S and 2023.A.00037.S. ALMA is a partnership of ESO (representing its member states), NSF (USA) and NINS (Japan), together with NRC (Canada), MOST and ASIAA (Taiwan), and KASI (Republic of Korea), in cooperation with the Republic of Chile. The Joint ALMA Observatory is operated by ESO, AUI/NRAO and NAOJ. We are greatly appreciative to our ALMA program coordinator Megan Lewis and also Violette Impellizzeri for support with our ALMA program and Allegro, the European ALMA Regional Center node in the Netherlands. We are grateful to J. Witstok for providing the {\sc Cloudy} models from \citet{Witstok_2022} for the analysis conducted in this paper. We thank M. Kohandel and A. Pallottini for simulated data analysis. AF acknowledges support from the ERC Advanced Grant INTERSTELLAR H2020/740120 and support from the grant NSF PHY-2309135 to the Kavli Institute for Theoretical Physics (KITP).
 
\end{acknowledgments}


\bibliography{sample631}{}

\begin{thebibliography}{}
\expandafter\ifx\csname natexlab\endcsname\relax\def\natexlab#1{#1}\fi
\providecommand{\url}[1]{\href{#1}{#1}}
\providecommand{\dodoi}[1]{doi:~\href{http://doi.org/#1}{\nolinkurl{#1}}}
\providecommand{\doeprint}[1]{\href{http://ascl.net/#1}{\nolinkurl{http://ascl.net/#1}}}
\providecommand{\doarXiv}[1]{\href{https://arxiv.org/abs/#1}{\nolinkurl{https://arxiv.org/abs/#1}}}

\bibitem[{{Akins} {et~al.}(2022){Akins}, {Fujimoto}, {Finlator}, {Watson}, {Knudsen}, {Richard}, {Bakx}, {Hashimoto}, {Inoue}, {Matsuo}, {Micha{\l}owski}, \& {Tamura}}]{akins2022}
{Akins}, H.~B., {Fujimoto}, S., {Finlator}, K., {et~al.} 2022, \apj, 934, 64, \dodoi{10.3847/1538-4357/ac795b}

\bibitem[{{Algera} {et~al.}(2025){Algera}, {Rowland}, {Stefanon}, {Palla}, {Sommovigo}, {Inami}, {Bouwens}, {Aravena}, {Bowler}, {Dayal}, {De Looze}, {Ferrara}, {Fisher}, {Graziani}, {Gulis}, {Heintz}, {Hodge}, {van Leeuwen}, {Pallottini}, {Phillips}, {Schouws}, {Smit}, {Stark}, \& {van der Werf}}]{algera2025}
{Algera}, H., {Rowland}, L., {Stefanon}, M., {et~al.} 2025, arXiv e-prints, arXiv:2501.10508, \dodoi{10.48550/arXiv.2501.10508}

\bibitem[{{Algera} {et~al.}(2024){Algera}, {Inami}, {Sommovigo}, {Fudamoto}, {Schneider}, {Graziani}, {Dayal}, {Bouwens}, {Aravena}, {da Cunha}, {Ferrara}, {Hygate}, {van Leeuwen}, {De Looze}, {Palla}, {Pallottini}, {Smit}, {Stefanon}, {Topping}, \& {van der Werf}}]{algera2024}
{Algera}, H. S.~B., {Inami}, H., {Sommovigo}, L., {et~al.} 2024, \mnras, 527, 6867, \dodoi{10.1093/mnras/stad3111}

\bibitem[{{{\'A}lvarez-M{\'a}rquez} {et~al.}(2024){{\'A}lvarez-M{\'a}rquez}, {Crespo G{\'o}mez}, {Colina}, {Langeroodi}, {Marques-Chaves}, {Prieto-Jim{\'e}nez}, {Bik}, {Alonso-Herrero}, {Boogaard}, {Costantin}, {Garc{\'\i}a-Mar{\'\i}n}, {Gillman}, {Hjorth}, {Iani}, {Jermann}, {Labiano}, {Melinder}, {Meyer}, {{\"O}stlin}, {P{\'e}rez-Gonz{\'a}lez}, {Rinaldi}, {Walter}, {van der Werf}, \& {Wright}}]{AlvarezMarquez2024}
{{\'A}lvarez-M{\'a}rquez}, J., {Crespo G{\'o}mez}, A., {Colina}, L., {et~al.} 2024, arXiv e-prints, arXiv:2412.12826, \dodoi{10.48550/arXiv.2412.12826}

\bibitem[{{Arata} {et~al.}(2020){Arata}, {Yajima}, {Nagamine}, {Abe}, \& {Khochfar}}]{arata2020}
{Arata}, S., {Yajima}, H., {Nagamine}, K., {Abe}, M., \& {Khochfar}, S. 2020, \mnras, 498, 5541, \dodoi{10.1093/mnras/staa2809}

\bibitem[{{Aravena} {et~al.}(2024){Aravena}, {Heintz}, {Dessauges-Zavadsky}, {Oesch}, {Algera}, {Bouwens}, {da Cunha}, {Dayal}, {De Looze}, {Ferrara}, {Fudamoto}, {Gonzalez}, {Graziani}, {Hygate}, {Inami}, {Pallottini}, {Schneider}, {Schouws}, {Sommovigo}, {Topping}, {van der Werf}, \& {Palla}}]{aravena2024}
{Aravena}, M., {Heintz}, K., {Dessauges-Zavadsky}, M., {et~al.} 2024, \aap, 682, A24, \dodoi{10.1051/0004-6361/202347281}

\bibitem[{{Bakx} {et~al.}(2020){Bakx}, {Tamura}, {Hashimoto}, {Inoue}, {Lee}, {Mawatari}, {Ota}, {Umehata}, {Zackrisson}, {Hatsukade}, {Kohno}, {Matsuda}, {Matsuo}, {Okamoto}, {Shibuya}, {Shimizu}, {Taniguchi}, \& {Yoshida}}]{bakx2020}
{Bakx}, T. J.~L.~C., {Tamura}, Y., {Hashimoto}, T., {et~al.} 2020, \mnras, 493, 4294, \dodoi{10.1093/mnras/staa509}

\bibitem[{{Bakx} {et~al.}(2023){Bakx}, {Zavala}, {Mitsuhashi}, {Treu}, {Fontana}, {Tadaki}, {Casey}, {Castellano}, {Glazebrook}, {Hagimoto}, {Ikeda}, {Jones}, {Leethochawalit}, {Mason}, {Morishita}, {Nanayakkara}, {Pentericci}, {Roberts-Borsani}, {Santini}, {Serjeant}, {Tamura}, {Trenti}, \& {Vanzella}}]{bakx2023}
{Bakx}, T. J.~L.~C., {Zavala}, J.~A., {Mitsuhashi}, I., {et~al.} 2023, \mnras, 519, 5076, \dodoi{10.1093/mnras/stac3723}

\bibitem[{{Bakx} {et~al.}(2024){Bakx}, {Algera}, {Venemans}, {Sommovigo}, {Fujimoto}, {Carniani}, {Hagimoto}, {Hashimoto}, {Inoue}, {Salak}, {Serjeant}, {Vallini}, {Eales}, {Ferrara}, {Fudamoto}, {Imamura}, {Inoue}, {Knudsen}, {Matsuo}, {Sugahara}, {Tamura}, {Taniguchi}, \& {Yamanaka}}]{bakx2024}
{Bakx}, T. J.~L.~C., {Algera}, H. S.~B., {Venemans}, B., {et~al.} 2024, \mnras, 532, 2270, \dodoi{10.1093/mnras/stae1613}

\bibitem[{{Binggeli} {et~al.}(2021){Binggeli}, {Inoue}, {Hashimoto}, {Toribio}, {Zackrisson}, {Ramstedt}, {Mawatari}, {Harikane}, {Matsuo}, {Okamoto}, {Ota}, {Shimizu}, {Tamura}, {Taniguchi}, \& {Umehata}}]{Binggeli2021}
{Binggeli}, C., {Inoue}, A.~K., {Hashimoto}, T., {et~al.} 2021, \aap, 646, A26, \dodoi{10.1051/0004-6361/202038180}

\bibitem[{{Bunker} {et~al.}(2023){Bunker}, {Saxena}, {Cameron}, {Willott}, {Curtis-Lake}, {Jakobsen}, {Carniani}, {Smit}, {Maiolino}, {Witstok}, {Curti}, {D'Eugenio}, {Jones}, {Ferruit}, {Arribas}, {Charlot}, {Chevallard}, {Giardino}, {de Graaff}, {Looser}, {L{\"u}tzgendorf}, {Maseda}, {Rawle}, {Rix}, {Del Pino}, {Alberts}, {Egami}, {Eisenstein}, {Endsley}, {Hainline}, {Hausen}, {Johnson}, {Rieke}, {Rieke}, {Robertson}, {Shivaei}, {Stark}, {Sun}, {Tacchella}, {Tang}, {Williams}, {Willmer}, {Baker}, {Baum}, {Bhatawdekar}, {Bowler}, {Boyett}, {Chen}, {Circosta}, {Helton}, {Ji}, {Kumari}, {Lyu}, {Nelson}, {Parlanti}, {Perna}, {Sandles}, {Scholtz}, {Suess}, {Topping}, {{\"U}bler}, {Wallace}, \& {Whitler}}]{bunker2023}
{Bunker}, A.~J., {Saxena}, A., {Cameron}, A.~J., {et~al.} 2023, \aap, 677, A88, \dodoi{10.1051/0004-6361/202346159}

\bibitem[{{Calabr{\`o}} {et~al.}(2024){Calabr{\`o}}, {Castellano}, {Zavala}, {Pentericci}, {Arrabal Haro}, {Bakx}, {Burgarella}, {Casey}, {Dickinson}, {Finkelstein}, {Fontana}, {Llerena}, {Mascia}, {Merlin}, {Mitsuhashi}, {Napolitano}, {Paris}, {P{\'e}rez-Gonz{\'a}lez}, {Roberts-Borsani}, {Santini}, {Treu}, \& {Vanzella}}]{calabro2024}
{Calabr{\`o}}, A., {Castellano}, M., {Zavala}, J.~A., {et~al.} 2024, \apj, 975, 245, \dodoi{10.3847/1538-4357/ad7602}

\bibitem[{{Carniani} {et~al.}(2017){Carniani}, {Maiolino}, {Pallottini}, {Vallini}, {Pentericci}, {Ferrara}, {Castellano}, {Vanzella}, {Grazian}, {Gallerani}, {Santini}, {Wagg}, \& {Fontana}}]{carniani2017}
{Carniani}, S., {Maiolino}, R., {Pallottini}, A., {et~al.} 2017, \aap, 605, A42, \dodoi{10.1051/0004-6361/201630366}

\bibitem[{{Carniani} {et~al.}(2020){Carniani}, {Ferrara}, {Maiolino}, {Castellano}, {Gallerani}, {Fontana}, {Kohandel}, {Lupi}, {Pallottini}, {Pentericci}, {Vallini}, \& {Vanzella}}]{carniani2020}
{Carniani}, S., {Ferrara}, A., {Maiolino}, R., {et~al.} 2020, \mnras, 499, 5136, \dodoi{10.1093/mnras/staa3178}

\bibitem[{{Carniani} {et~al.}(2024{\natexlab{a}}){Carniani}, {D'Eugenio}, {Ji}, {Parlanti}, {Scholtz}, {Sun}, {Venturi}, {Bakx}, {Curti}, {Maiolino}, {Tacchella}, {Zavala}, {Hainline}, {Witstok}, {Johnson}, {Alberts}, {Bunker}, {Charlot}, {Eisenstein}, {Helton}, {Jakobsen}, {Kumari}, {Robertson}, {Saxena}, {{\"U}bler}, {Williams}, {Willmer}, \& {Willott}}]{carniani2024_oiii}
{Carniani}, S., {D'Eugenio}, F., {Ji}, X., {et~al.} 2024{\natexlab{a}}, arXiv e-prints, arXiv:2409.20533, \dodoi{10.48550/arXiv.2409.20533}

\bibitem[{{Carniani} {et~al.}(2024{\natexlab{b}}){Carniani}, {Hainline}, {D'Eugenio}, {Eisenstein}, {Jakobsen}, {Witstok}, {Johnson}, {Chevallard}, {Maiolino}, {Helton}, {Willott}, {Robertson}, {Alberts}, {Arribas}, {Baker}, {Bhatawdekar}, {Boyett}, {Bunker}, {Cameron}, {Cargile}, {Charlot}, {Curti}, {Curtis-Lake}, {Egami}, {Giardino}, {Isaak}, {Ji}, {Jones}, {Kumari}, {Maseda}, {Parlanti}, {P{\'e}rez-Gonz{\'a}lez}, {Rawle}, {Rieke}, {Rieke}, {Del Pino}, {Saxena}, {Scholtz}, {Smit}, {Sun}, {Tacchella}, {{\"U}bler}, {Venturi}, {Williams}, \& {Willmer}}]{carniani2024}
{Carniani}, S., {Hainline}, K., {D'Eugenio}, F., {et~al.} 2024{\natexlab{b}}, \nat, 633, 318, \dodoi{10.1038/s41586-024-07860-9}

\bibitem[{{Casavecchia} {et~al.}(2024){Casavecchia}, {Maio}, {P{\'e}roux}, \& {Ciardi}}]{casavecchia2024}
{Casavecchia}, B., {Maio}, U., {P{\'e}roux}, C., \& {Ciardi}, B. 2024, \aap, 689, A106, \dodoi{10.1051/0004-6361/202450332}

\bibitem[{{Castellano} {et~al.}(2024){Castellano}, {Napolitano}, {Fontana}, {Roberts-Borsani}, {Treu}, {Vanzella}, {Zavala}, {Arrabal Haro}, {Calabr{\`o}}, {Llerena}, {Mascia}, {Merlin}, {Paris}, {Pentericci}, {Santini}, {Bakx}, {Bergamini}, {Cupani}, {Dickinson}, {Filippenko}, {Glazebrook}, {Grillo}, {Kelly}, {Malkan}, {Mason}, {Morishita}, {Nanayakkara}, {Rosati}, {Sani}, {Wang}, \& {Yoon}}]{castellano2024}
{Castellano}, M., {Napolitano}, L., {Fontana}, A., {et~al.} 2024, \apj, 972, 143, \dodoi{10.3847/1538-4357/ad5f88}

\bibitem[{{Chabrier}(2003)}]{chabrier2003}
{Chabrier}, G. 2003, \pasp, 115, 763, \dodoi{10.1086/376392}

\bibitem[{{Cormier} {et~al.}(2019){Cormier}, {Abel}, {Hony}, {Lebouteiller}, {Madden}, {Polles}, {Galliano}, {De Looze}, {Galametz}, \& {Lambert-Huyghe}}]{cormier2019}
{Cormier}, D., {Abel}, N.~P., {Hony}, S., {et~al.} 2019, \aap, 626, A23, \dodoi{10.1051/0004-6361/201834457}

\bibitem[{{Curtis-Lake} {et~al.}(2023){Curtis-Lake}, {Carniani}, {Cameron}, {Charlot}, {Jakobsen}, {Maiolino}, {Bunker}, {Witstok}, {Smit}, {Chevallard}, {Willott}, {Ferruit}, {Arribas}, {Bonaventura}, {Curti}, {D'Eugenio}, {Franx}, {Giardino}, {Looser}, {L{\"u}tzgendorf}, {Maseda}, {Rawle}, {Rix}, {Rodr{\'\i}guez del Pino}, {{\"U}bler}, {Sirianni}, {Dressler}, {Egami}, {Eisenstein}, {Endsley}, {Hainline}, {Hausen}, {Johnson}, {Rieke}, {Robertson}, {Shivaei}, {Stark}, {Tacchella}, {Williams}, {Willmer}, {Bhatawdekar}, {Bowler}, {Boyett}, {Chen}, {de Graaff}, {Helton}, {Hviding}, {Jones}, {Kumari}, {Lyu}, {Nelson}, {Perna}, {Sandles}, {Saxena}, {Suess}, {Sun}, {Topping}, {Wallace}, \& {Whitler}}]{Curtis-Lake_2022}
{Curtis-Lake}, E., {Carniani}, S., {Cameron}, A., {et~al.} 2023, Nature Astronomy, 7, 622, \dodoi{10.1038/s41550-023-01918-w}

\bibitem[{{De Looze} {et~al.}(2014){De Looze}, {Cormier}, {Lebouteiller}, {Madden}, {Baes}, {Bendo}, {Boquien}, {Boselli}, {Clements}, {Cortese}, {Cooray}, {Galametz}, {Galliano}, {Graci{\'a}-Carpio}, {Isaak}, {Karczewski}, {Parkin}, {Pellegrini}, {R{\'e}my-Ruyer}, {Spinoglio}, {Smith}, \& {Sturm}}]{delooze2014}
{De Looze}, I., {Cormier}, D., {Lebouteiller}, V., {et~al.} 2014, \aap, 568, A62, \dodoi{10.1051/0004-6361/201322489}

\bibitem[{{De Vis} {et~al.}(2019){De Vis}, {Jones}, {Viaene}, {Casasola}, {Clark}, {Baes}, {Bianchi}, {Cassara}, {Davies}, {De Looze}, {Galametz}, {Galliano}, {Lianou}, {Madden}, {Manilla-Robles}, {Mosenkov}, {Nersesian}, {Roychowdhury}, {Xilouris}, \& {Ysard}}]{devis2019}
{De Vis}, P., {Jones}, A., {Viaene}, S., {et~al.} 2019, \aap, 623, A5, \dodoi{10.1051/0004-6361/201834444}

\bibitem[{{Dessauges-Zavadsky} {et~al.}(2020){Dessauges-Zavadsky}, {Ginolfi}, {Pozzi}, {B{\'e}thermin}, {Le F{\`e}vre}, {Fujimoto}, {Silverman}, {Jones}, {Vallini}, {Schaerer}, {Faisst}, {Khusanova}, {Fudamoto}, {Cassata}, {Loiacono}, {Capak}, {Yan}, {Amorin}, {Bardelli}, {Boquien}, {Cimatti}, {Gruppioni}, {Hathi}, {Ibar}, {Koekemoer}, {Lemaux}, {Narayanan}, {Oesch}, {Rodighiero}, {Romano}, {Talia}, {Toft}, {Vergani}, {Zamorani}, \& {Zucca}}]{dessauges-zavadsky2020}
{Dessauges-Zavadsky}, M., {Ginolfi}, M., {Pozzi}, F., {et~al.} 2020, \aap, 643, A5, \dodoi{10.1051/0004-6361/202038231}

\bibitem[{{Eisenstein} {et~al.}(2023{\natexlab{a}}){Eisenstein}, {Willott}, {Alberts}, {Arribas}, {Bonaventura}, {Bunker}, {Cameron}, {Carniani}, {Charlot}, {Curtis-Lake}, {D'Eugenio}, {Endsley}, {Ferruit}, {Giardino}, {Hainline}, {Hausen}, {Jakobsen}, {Johnson}, {Maiolino}, {Rieke}, {Rieke}, {Rix}, {Robertson}, {Stark}, {Tacchella}, {Williams}, {Willmer}, {Baker}, {Baum}, {Bhatawdekar}, {Boyett}, {Chen}, {Chevallard}, {Circosta}, {Curti}, {Danhaive}, {DeCoursey}, {de Graaff}, {Dressler}, {Egami}, {Helton}, {Hviding}, {Ji}, {Jones}, {Kumari}, {L{\"u}tzgendorf}, {Laseter}, {Looser}, {Lyu}, {Maseda}, {Nelson}, {Parlanti}, {Perna}, {Pusk{\'a}s}, {Rawle}, {Rodr{\'\i}guez Del Pino}, {Sandles}, {Saxena}, {Scholtz}, {Sharpe}, {Shivaei}, {Silcock}, {Simmonds}, {Skarbinski}, {Smit}, {Stone}, {Suess}, {Sun}, {Tang}, {Topping}, {{\"U}bler}, {Villanueva}, {Wallace}, {Whitler}, {Witstok}, \& {Woodrum}}]{Eisenstein_2023}
{Eisenstein}, D.~J., {Willott}, C., {Alberts}, S., {et~al.} 2023{\natexlab{a}}, arXiv e-prints, arXiv:2306.02465, \dodoi{10.48550/arXiv.2306.02465}

\bibitem[{{Eisenstein} {et~al.}(2023{\natexlab{b}}){Eisenstein}, {Johnson}, {Robertson}, {Tacchella}, {Hainline}, {Jakobsen}, {Maiolino}, {Bonaventura}, {Bunker}, {Cameron}, {Cargile}, {Curtis-Lake}, {Hausen}, {Pusk{\'a}s}, {Rieke}, {Sun}, {Willmer}, {Willott}, {Alberts}, {Arribas}, {Baker}, {Baum}, {Bhatawdekar}, {Carniani}, {Charlot}, {Chen}, {Chevallard}, {Curti}, {DeCoursey}, {D'Eugenio}, {de Graaff}, {Egami}, {Helton}, {Ji}, {Jones}, {Kumari}, {L{\"u}tzgendorf}, {Laseter}, {Looser}, {Lyu}, {Maseda}, {Nelson}, {Parlanti}, {Rauscher}, {Rawle}, {Rieke}, {Rix}, {Rujopakarn}, {Sandles}, {Saxena}, {Scholtz}, {Sharpe}, {Shivaei}, {Simmonds}, {Smit}, {Topping}, {{\"U}bler}, {Venturi}, {Williams}, {Witstok}, \& {Woodrum}}]{Eisenstein_2023_JOF}
{Eisenstein}, D.~J., {Johnson}, B.~D., {Robertson}, B., {et~al.} 2023{\natexlab{b}}, arXiv e-prints, arXiv:2310.12340, \dodoi{10.48550/arXiv.2310.12340}

\bibitem[{{Ferland} {et~al.}(2017){Ferland}, {Chatzikos}, {Guzm{\'a}n}, {Lykins}, {van Hoof}, {Williams}, {Abel}, {Badnell}, {Keenan}, {Porter}, \& {Stancil}}]{ferland2017}
{Ferland}, G.~J., {Chatzikos}, M., {Guzm{\'a}n}, F., {et~al.} 2017, \rmxaa, 53, 385.
\newblock \doarXiv{1705.10877}

\bibitem[{{Ferrara}(2024)}]{Ferrara2024a}
{Ferrara}, A. 2024, \aap, 689, A310, \dodoi{10.1051/0004-6361/202450944}

\bibitem[{Ferrara {et~al.}(2024)Ferrara, Carniani, di~Mascia, Bouwens, Oesch, \& Schouws}]{Ferrara2024b}
Ferrara, A., Carniani, S., di~Mascia, F., {et~al.} 2024.
\newblock \doarXiv{2409.17223}

\bibitem[{{Ferrara} {et~al.}(2019){Ferrara}, {Vallini}, {Pallottini}, {Gallerani}, {Carniani}, {Kohandel}, {Decataldo}, \& {Behrens}}]{ferrara2019}
{Ferrara}, A., {Vallini}, L., {Pallottini}, A., {et~al.} 2019, \mnras, 489, 1, \dodoi{10.1093/mnras/stz2031}

\bibitem[{{Fujimoto} {et~al.}(2023){Fujimoto}, {Finkelstein}, {Burgarella}, {Carilli}, {Buat}, {Casey}, {Ciesla}, {Tacchella}, {Zavala}, {Brammer}, {Fudamoto}, {Ouchi}, {Valentino}, {Cooper}, {Dickinson}, {Franco}, {Giavalisco}, {Hutchison}, {Kartaltepe}, {Koekemoer}, {Kojima}, {Larson}, {Murphy}, {Papovich}, {P{\'e}rez-Gonz{\'a}lez}, {Somerville}, {Yoon}, {Wilkins}, {Akins}, {Amor{\'\i}n}, {Arrabal Haro}, {Bagley}, {Chworowsky}, {Cleri}, {Cooper}, {Costantin}, {Daddi}, {Ferguson}, {Grogin}, {Jim{\'e}nez-Andrade}, {Juneau}, {Kirkpatrick}, {Kocevski}, {Le Bail}, {Long}, {Lucas}, {Magnelli}, {McKinney}, {Rose}, {Seill{\'e}}, {Simons}, {Weiner}, \& {Yung}}]{fujimoto2023_highzalma}
{Fujimoto}, S., {Finkelstein}, S.~L., {Burgarella}, D., {et~al.} 2023, \apj, 955, 130, \dodoi{10.3847/1538-4357/aceb67}

\bibitem[{{Fujimoto} {et~al.}(2024){Fujimoto}, {Ouchi}, {Nakajima}, {Harikane}, {Isobe}, {Brammer}, {Oguri}, {Gim{\'e}nez-Arteaga}, {Heintz}, {Kokorev}, {Bauer}, {Ferrara}, {Kojima}, {Lagos}, {Laura}, {Schaerer}, {Shimasaku}, {Hatsukade}, {Kohno}, {Sun}, {Valentino}, {Watson}, {Fudamoto}, {Inoue}, {Gonz{\'a}lez-L{\'o}pez}, {Koekemoer}, {Knudsen}, {Lee}, {Magdis}, {Richard}, {Strait}, {Sugahara}, {Tamura}, {Toft}, {Umehata}, \& {Walth}}]{fujimoto2024_s04950}
{Fujimoto}, S., {Ouchi}, M., {Nakajima}, K., {et~al.} 2024, \apj, 964, 146, \dodoi{10.3847/1538-4357/ad235c}

\bibitem[{{Glazer} {et~al.}(2024){Glazer}, {Brad{\u{a}}c}, {Sanders}, {Fujimoto}, {Bolan}, {Ferrara}, {Strait}, {Jones}, {Lemaux}, {Vallini}, \& {Ryan}}]{glazer2024}
{Glazer}, K., {Brad{\u{a}}c}, M., {Sanders}, R.~L., {et~al.} 2024, \mnras, 531, 945, \dodoi{10.1093/mnras/stae1178}

\bibitem[{{Gong} {et~al.}(2012){Gong}, {Cooray}, {Silva}, {Santos}, {Bock}, {Bradford}, \& {Zemcov}}]{gong2012}
{Gong}, Y., {Cooray}, A., {Silva}, M., {et~al.} 2012, \apj, 745, 49, \dodoi{10.1088/0004-637X/745/1/49}

\bibitem[{{Gullberg} {et~al.}(2015){Gullberg}, {De Breuck}, {Vieira}, {Wei{\ss}}, {Aguirre}, {Aravena}, {B{\'e}thermin}, {Bradford}, {Bothwell}, {Carlstrom}, {Chapman}, {Fassnacht}, {Gonzalez}, {Greve}, {Hezaveh}, {Holzapfel}, {Husband}, {Ma}, {Malkan}, {Marrone}, {Menten}, {Murphy}, {Reichardt}, {Spilker}, {Stark}, {Strandet}, \& {Welikala}}]{gullberg2015}
{Gullberg}, B., {De Breuck}, C., {Vieira}, J.~D., {et~al.} 2015, \mnras, 449, 2883, \dodoi{10.1093/mnras/stv372}

\bibitem[{{Harikane} {et~al.}(2020){Harikane}, {Ouchi}, {Inoue}, {Matsuoka}, {Tamura}, {Bakx}, {Fujimoto}, {Moriwaki}, {Ono}, {Nagao}, {Tadaki}, {Kojima}, {Shibuya}, {Egami}, {Ferrara}, {Gallerani}, {Hashimoto}, {Kohno}, {Matsuda}, {Matsuo}, {Pallottini}, {Sugahara}, \& {Vallini}}]{harikane2020}
{Harikane}, Y., {Ouchi}, M., {Inoue}, A.~K., {et~al.} 2020, \apj, 896, 93, \dodoi{10.3847/1538-4357/ab94bd}

\bibitem[{{Harikane} {et~al.}(2023){Harikane}, {Ouchi}, {Oguri}, {Ono}, {Nakajima}, {Isobe}, {Umeda}, {Mawatari}, \& {Zhang}}]{harikane2022}
{Harikane}, Y., {Ouchi}, M., {Oguri}, M., {et~al.} 2023, \apjs, 265, 5, \dodoi{10.3847/1538-4365/acaaa9}

\bibitem[{{Hashimoto} {et~al.}(2019){Hashimoto}, {Inoue}, {Mawatari}, {Tamura}, {Matsuo}, {Furusawa}, {Harikane}, {Shibuya}, {Knudsen}, {Kohno}, {Ono}, {Zackrisson}, {Okamoto}, {Kashikawa}, {Oesch}, {Ouchi}, {Ota}, {Shimizu}, {Taniguchi}, {Umehata}, \& {Watson}}]{hashimoto2019}
{Hashimoto}, T., {Inoue}, A.~K., {Mawatari}, K., {et~al.} 2019, \pasj, 71, 71, \dodoi{10.1093/pasj/psz049}

\bibitem[{Helton {et~al.}(2024)Helton, Rieke, Alberts, Wu, Eisenstein, Hainline, Carniani, Ji, Baker, Bhatawdekar, Bunker, Cargile, Charlot, Chevallard, D'Eugenio, Egami, Johnson, Jones, Lyu, Maiolino, Pérez-González, Rieke, Robertson, Saxena, Scholtz, Shivaei, Sun, Tacchella, Whitler, Williams, Willmer, Willott, Witstok, \& Zhu}]{helton2024}
Helton, J.~M., Rieke, G.~H., Alberts, S., {et~al.} 2024.
\newblock \doarXiv{2405.18462}

\bibitem[{{Herrera-Camus} {et~al.}(2015){Herrera-Camus}, {Bolatto}, {Wolfire}, {Smith}, {Croxall}, {Kennicutt}, {Calzetti}, {Helou}, {Walter}, {Leroy}, {Draine}, {Brandl}, {Armus}, {Sandstrom}, {Dale}, {Aniano}, {Meidt}, {Boquien}, {Hunt}, {Galametz}, {Tabatabaei}, {Murphy}, {Appleton}, {Roussel}, {Engelbracht}, \& {Beirao}}]{herrera-camus2015}
{Herrera-Camus}, R., {Bolatto}, A.~D., {Wolfire}, M.~G., {et~al.} 2015, \apj, 800, 1, \dodoi{10.1088/0004-637X/800/1/1}

\bibitem[{Hunter {et~al.}(2023)Hunter, Indebetouw, Brogan, Berry, Chang, Francke, Geers, Gómez, Hibbard, Humphreys, Kent, Kepley, Kunneriath, Lipnicky, Loomis, Mason, Masters, Maud, Muders, Sabater, Sugimoto, Szűcs, Vasiliev, Videla, Villard, Williams, Xue, \& Yoon}]{Hunter_2023}
Hunter, T.~R., Indebetouw, R., Brogan, C.~L., {et~al.} 2023, Publications of the Astronomical Society of the Pacific, 135, 074501, \dodoi{10.1088/1538-3873/ace216}

\bibitem[{{Inoue} {et~al.}(2016){Inoue}, {Tamura}, {Matsuo}, {Mawatari}, {Shimizu}, {Shibuya}, {Ota}, {Yoshida}, {Zackrisson}, {Kashikawa}, {Kohno}, {Umehata}, {Hatsukade}, {Iye}, {Matsuda}, {Okamoto}, \& {Yamaguchi}}]{inoue2016}
{Inoue}, A.~K., {Tamura}, Y., {Matsuo}, H., {et~al.} 2016, Science, 352, 1559, \dodoi{10.1126/science.aaf0714}

\bibitem[{{Isobe} {et~al.}(2023){Isobe}, {Ouchi}, {Nakajima}, {Harikane}, {Ono}, {Xu}, {Zhang}, \& {Umeda}}]{isobe2023}
{Isobe}, Y., {Ouchi}, M., {Nakajima}, K., {et~al.} 2023, \apj, 956, 139, \dodoi{10.3847/1538-4357/acf376}

\bibitem[{{Kaasinen} {et~al.}(2023){Kaasinen}, {van Marrewijk}, {Popping}, {Ginolfi}, {Di Mascolo}, {Mroczkowski}, {Concas}, {Di Cesare}, {Killi}, \& {Langan}}]{kaasinen2023}
{Kaasinen}, M., {van Marrewijk}, J., {Popping}, G., {et~al.} 2023, \aap, 671, A29, \dodoi{10.1051/0004-6361/202245093}

\bibitem[{{Katz} {et~al.}(2022){Katz}, {Rosdahl}, {Kimm}, {Garel}, {Blaizot}, {Haehnelt}, {Michel-Dansac}, {Martin-Alvarez}, {Devriendt}, {Slyz}, {Teyssier}, {Ocvirk}, {Laporte}, \& {Ellis}}]{katz2022}
{Katz}, H., {Rosdahl}, J., {Kimm}, T., {et~al.} 2022, \mnras, 510, 5603, \dodoi{10.1093/mnras/stac028}

\bibitem[{{Kepley} {et~al.}(2020){Kepley}, {Tsutsumi}, {Brogan}, {Indebetouw}, {Yoon}, {Mason}, \& {Donovan Meyer}}]{kepley2020}
{Kepley}, A.~A., {Tsutsumi}, T., {Brogan}, C.~L., {et~al.} 2020, \pasp, 132, 024505, \dodoi{10.1088/1538-3873/ab5e14}

\bibitem[{{Kohandel} {et~al.}(2019){Kohandel}, {Pallottini}, {Ferrara}, {Zanella}, {Behrens}, {Carniani}, {Gallerani}, \& {Vallini}}]{kohandel2019}
{Kohandel}, M., {Pallottini}, A., {Ferrara}, A., {et~al.} 2019, \mnras, 487, 3007, \dodoi{10.1093/mnras/stz1486}

\bibitem[{{Kumari} {et~al.}(2024){Kumari}, {Smit}, {Leitherer}, {Witstok}, {Irwin}, {Sirianni}, \& {Aloisi}}]{Kumari2024a}
{Kumari}, N., {Smit}, R., {Leitherer}, C., {et~al.} 2024, \mnras, 529, 781, \dodoi{10.1093/mnras/stae252}

\bibitem[{{Laporte} {et~al.}(2019){Laporte}, {Katz}, {Ellis}, {Lagache}, {Bauer}, {Boone}, {Inoue}, {Hashimoto}, {Matsuo}, {Mawatari}, \& {Tamura}}]{laporte2019}
{Laporte}, N., {Katz}, H., {Ellis}, R.~S., {et~al.} 2019, \mnras, 487, L81, \dodoi{10.1093/mnrasl/slz094}

\bibitem[{{Nakazato} {et~al.}(2023){Nakazato}, {Yoshida}, \& {Ceverino}}]{nakazato2023}
{Nakazato}, Y., {Yoshida}, N., \& {Ceverino}, D. 2023, \apj, 953, 140, \dodoi{10.3847/1538-4357/ace25a}

\bibitem[{{Novak} {et~al.}(2019){Novak}, {Ba{\~n}ados}, {Decarli}, {Walter}, {Venemans}, {Neeleman}, {Farina}, {Mazzucchelli}, {Carilli}, {Fan}, {Rix}, \& {Wang}}]{novak2019}
{Novak}, M., {Ba{\~n}ados}, E., {Decarli}, R., {et~al.} 2019, \apj, 881, 63, \dodoi{10.3847/1538-4357/ab2beb}

\bibitem[{{Nyhagen} {et~al.}(2024){Nyhagen}, {Schimek}, {Cicone}, {Decataldo}, \& {Shen}}]{nyhagen2024}
{Nyhagen}, C.~T., {Schimek}, A., {Cicone}, C., {Decataldo}, D., \& {Shen}, S. 2024, arXiv e-prints, arXiv:2410.18471, \dodoi{10.48550/arXiv.2410.18471}

\bibitem[{Oesch {et~al.}(2016)Oesch, Brammer, Dokkum, Illingworth, Bouwens, Labbé, Franx, Momcheva, Ashby, Fazio, Gonzalez, Holden, Magee, Skelton, Smit, Spitler, Trenti, \& Willner}]{Oesch_2016}
Oesch, P.~A., Brammer, G., Dokkum, P. G.~v., {et~al.} 2016, The Astrophysical Journal, 819, 129, \dodoi{10.3847/0004-637x/819/2/129}

\bibitem[{{Oesch} {et~al.}(2023){Oesch}, {Brammer}, {Naidu}, {Bouwens}, {Chisholm}, {Illingworth}, {Matthee}, {Nelson}, {Qin}, {Reddy}, {Shapley}, {Shivaei}, {van Dokkum}, {Weibel}, {Whitaker}, {Wuyts}, {Covelo-Paz}, {Endsley}, {Fudamoto}, {Giovinazzo}, {Herard-Demanche}, {Kerutt}, {Kramarenko}, {Labbe}, {Leonova}, {Lin}, {Magee}, {Marchesini}, {Maseda}, {Mason}, {Matharu}, {Meyer}, {Neufeld}, {Prieto Lyon}, {Schaerer}, {Sharma}, {Shuntov}, {Smit}, {Stefanon}, {Wyithe}, \& {Xiao}}]{Oesch_2023}
{Oesch}, P.~A., {Brammer}, G., {Naidu}, R.~P., {et~al.} 2023, \mnras, 525, 2864, \dodoi{10.1093/mnras/stad2411}

\bibitem[{{Oke} \& {Gunn}(1983)}]{oke_gunn_1983ApJ...266..713O}
{Oke}, J.~B., \& {Gunn}, J.~E. 1983, \apj, 266, 713, \dodoi{10.1086/160817}

\bibitem[{Popping(2023)}]{Popping_2023}
Popping, G. 2023, Astronomy \& Astrophysics, 669, L8, \dodoi{10.1051/0004-6361/202244831}

\bibitem[{{Reddy} {et~al.}(2023){Reddy}, {Topping}, {Sanders}, {Shapley}, \& {Brammer}}]{reddy2023}
{Reddy}, N.~A., {Topping}, M.~W., {Sanders}, R.~L., {Shapley}, A.~E., \& {Brammer}, G. 2023, \apj, 952, 167, \dodoi{10.3847/1538-4357/acd754}

\bibitem[{{Robertson}(2022)}]{Robertson2022}
{Robertson}, B.~E. 2022, \araa, 60, 121, \dodoi{10.1146/annurev-astro-120221-044656}

\bibitem[{{Schaerer} {et~al.}(2020){Schaerer}, {Ginolfi}, {B{\'e}thermin}, {Fudamoto}, {Oesch}, {Le F{\`e}vre}, {Faisst}, {Capak}, {Cassata}, {Silverman}, {Yan}, {Jones}, {Amorin}, {Bardelli}, {Boquien}, {Cimatti}, {Dessauges-Zavadsky}, {Giavalisco}, {Hathi}, {Fujimoto}, {Ibar}, {Koekemoer}, {Lagache}, {Lemaux}, {Loiacono}, {Maiolino}, {Narayanan}, {Morselli}, {M{\'e}ndez-Hern{\`a}ndez}, {Pozzi}, {Riechers}, {Talia}, {Toft}, {Vallini}, {Vergani}, {Zamorani}, \& {Zucca}}]{schaerer2020}
{Schaerer}, D., {Ginolfi}, M., {B{\'e}thermin}, M., {et~al.} 2020, \aap, 643, A3, \dodoi{10.1051/0004-6361/202037617}

\bibitem[{{Schouws} {et~al.}(2024){Schouws}, {Bouwens}, {Ormerod}, {Smit}, {Algera}, {Sommovigo}, {Hodge}, {Ferrara}, {Oesch}, {Rowland}, {van Leeuwen}, {Stefanon}, {Herard-Demanche}, {Fudamoto}, {R{\"o}ttgering}, \& {van der Werf}}]{schouws2024}
{Schouws}, S., {Bouwens}, R.~J., {Ormerod}, K., {et~al.} 2024, arXiv e-prints, arXiv:2409.20549, \dodoi{10.48550/arXiv.2409.20549}

\bibitem[{{Solomon} {et~al.}(1992){Solomon}, {Downes}, \& {Radford}}]{Solomon_1992}
{Solomon}, P.~M., {Downes}, D., \& {Radford}, S.~J.~E. 1992, \apjl, 387, L55, \dodoi{10.1086/186304}

\bibitem[{{Tang} {et~al.}(2023){Tang}, {Stark}, {Chen}, {Mason}, {Topping}, {Endsley}, {Senchyna}, {Plat}, {Lu}, {Whitler}, {Robertson}, \& {Charlot}}]{tang2023}
{Tang}, M., {Stark}, D.~P., {Chen}, Z., {et~al.} 2023, \mnras, 526, 1657, \dodoi{10.1093/mnras/stad2763}

\bibitem[{{Thompson} {et~al.}(2017){Thompson}, {Moran}, \& {Swenson}}]{thompson2017}
{Thompson}, A.~R., {Moran}, J.~M., \& {Swenson}, George~W., J. 2017, {Interferometry and Synthesis in Radio Astronomy, 3rd Edition}, \dodoi{10.1007/978-3-319-44431-4}

\bibitem[{{Witstok} {et~al.}(2022){Witstok}, {Smit}, {Maiolino}, {Kumari}, {Aravena}, {Boogaard}, {Bouwens}, {Carniani}, {Hodge}, {Jones}, {Stefanon}, {van der Werf}, \& {Schouws}}]{Witstok_2022}
{Witstok}, J., {Smit}, R., {Maiolino}, R., {et~al.} 2022, \mnras, 515, 1751, \dodoi{10.1093/mnras/stac1905}

\bibitem[{Yoon {et~al.}(2023)Yoon, Carilli, Fujimoto, Castellano, Merlin, Santini, Yun, Murphy, Jung, Casey, Finkelstein, Papovich, Fontana, Treu, \& Letai}]{Yoon_2023}
Yoon, I., Carilli, C.~L., Fujimoto, S., {et~al.} 2023, The Astrophysical Journal, 950, 61, \dodoi{10.3847/1538-4357/acc94d}

\bibitem[{{Zanella} {et~al.}(2018){Zanella}, {Daddi}, {Magdis}, {Diaz Santos}, {Cormier}, {Liu}, {Cibinel}, {Gobat}, {Dickinson}, {Sargent}, {Popping}, {Madden}, {Bethermin}, {Hughes}, {Valentino}, {Rujopakarn}, {Pannella}, {Bournaud}, {Walter}, {Wang}, {Elbaz}, \& {Coogan}}]{zanella2018}
{Zanella}, A., {Daddi}, E., {Magdis}, G., {et~al.} 2018, \mnras, 481, 1976, \dodoi{10.1093/mnras/sty2394}

\bibitem[{{Zavala} {et~al.}(2024{\natexlab{a}}){Zavala}, {Castellano}, {Akins}, {Bakx}, {Burgarella}, {Casey}, {Ch{\'a}vez Ortiz}, {Dickinson}, {Finkelstein}, {Mitsuhashi}, {Nakajima}, {P{\'e}rez-Gonz{\'a}lez}, {Arrabal Haro}, {Buat}, {Backhaus}, {Calabr{\`o}}, {Cleri}, {Fern{\'a}ndez-Arenas}, {Fontana}, {Franco}, {Giavalisco}, {Grogin}, {Hathi}, {Hirschmann}, {Ikeda}, {Jung}, {Kartaltepe}, {Koekemoer}, {Larson}, {McKinney}, {Papovich}, {Saito}, {Santini}, {Terlevich}, {Terlevich}, {Treu}, \& {Yung}}]{zavala2024}
{Zavala}, J.~A., {Castellano}, M., {Akins}, H.~B., {et~al.} 2024{\natexlab{a}}, arXiv e-prints, arXiv:2403.10491, \dodoi{10.48550/arXiv.2403.10491}

\bibitem[{{Zavala} {et~al.}(2024{\natexlab{b}}){Zavala}, {Bakx}, {Mitsuhashi}, {Castellano}, {Calabro}, {Akins}, {Buat}, {Casey}, {Fernandez-Arenas}, {Franco}, {Fontana}, {Hatsukade}, {Ho}, {Ikeda}, {Kartaltepe}, {Koekemoer}, {McKinney}, {Napolitano}, {P{\'e}rez-Gonz{\'a}lez}, {Santini}, {Serjeant}, {Terlevich}, {Terlevich}, \& {Yung}}]{zavala2024_alma}
{Zavala}, J.~A., {Bakx}, T., {Mitsuhashi}, I., {et~al.} 2024{\natexlab{b}}, \apjl, 977, L9, \dodoi{10.3847/2041-8213/ad8f38}

\end{thebibliography}
\bibliographystyle{aasjournal}

\end{document}